\documentstyle[12pt]{article}

\newcommand{\beq}{\begin{equation}}
\newcommand{\eeq}{\end{equation}}
\newcommand{\bea}{\begin{eqnarray}}
\newcommand{\eea}{\end{eqnarray}}
\def\half{{\textstyle{1\over 2}}}
\def\quarter{{\textstyle{1\over 4}}}
\def\as{\alpha_s}
\def\asb{{\overline\alpha}_s}
\def\asnot{\alpha_{0}}
\def\df{\Delta\phi}
\def\dy{\Delta y}
\def\pti{p_{T1}}
\def\ptii{p_{T2}}
\def\vpti{\vec{p}_{T1}}
\def\vptii{\vec{p}_{T2}}
\def\ptisq{p_{T1}^2}
\def\ptiisq{p_{T2}^2}
\def\tf{{\tilde f}}
\def\d{{\rm d}}

\def\Dzero{D$\emptyset$}
\def\pp{{\rm p\bar{p}}}
\def\Re{{\rm Re}}
\def\GeV{{\rm GeV}}
\def\MeV{{\rm MeV}}

\def\cO{{\cal O}}
\def\lapprox{\lower .7ex\hbox{$\;\stackrel{\textstyle <}{\sim}\;$}}
\def\gapprox{\lower .7ex\hbox{$\;\stackrel{\textstyle >}{\sim}\;$}}
\def\figi{1}
\def\figii{2}
\def\figiii{3}
\def\figiv{4}
\def\figv{5}

\begin{document}
\titlepage
\begin{flushright}
{DTP/97/48}\\
{UR-1499}\\
{June 1997}\\
{hep-ph/9706529}\\
\end{flushright}
\begin{center}
\vspace*{2cm}
{\Large {\bf Dijet Production at Hadron--Hadron Colliders \\ [2mm]
in the BFKL Approach}} \\

\vspace*{1.5cm}
Lynne H.\ Orr$^a$ and W.\ J.\ Stirling$^{b}$ \\

\vspace*{0.5cm}
$^a \; $ {\it Department of Physics and Astronomy, University of Rochester,
Rochester, NY~14627-0171, USA}\\

$^b \; $ {\it Departments of Physics and  Mathematical Sciences, 
University of Durham,
Durham, DH1~3LE, UK}\\

\end{center}

\vspace*{4cm}
\begin{abstract}
The production in high-energy hadron collisions
of a pair of jets with large rapidity separation is studied in an improved BFKL
formalism. By recasting the analytic solution of the BFKL equation as
an explicit order-by-order sum over emitted gluons, the effects of phase
space constraints and the running coupling are studied. Particular attention
is paid to the azimuthal angle decorrelation of the jet pair. The inclusion
of sub-leading effects  significantly improves the agreement
between the theoretical predictions and recent preliminary 
measurements from the
\Dzero\ collaboration.
\end{abstract}

\newpage
\section{Introduction}
Fixed-order, renormalization-group-improved 
QCD perturbation theory has been remarkably successful
in describing jet physics at high-energy colliders.
For example, the bulk of the inclusive jet distribution
at the Tevatron $\pp$ collider is well described by folding
subprocess cross sections calculated at NLO with parton distributions
extracted from deep inelastic scattering. However there are
certain situations where such an approach may be expected to fail.
If in the production of a pair of large $E_T$ jets 
the separation in rapidity $\dy$ of the jets becomes large,
then higher-order perturbative corrections become more and more
important. Essentially, for $\dy \gg 1$
the subprocess cross section has an expansion in powers of
$\alpha_s\dy$ rather than $\alpha_s$. 

Dijet production with a large rapidity separation is an example
of a `two-large-scale' process in perturbative QCD, where
large logarithms, in this case $\ln(\hat{s}/E_T^2) \sim 
\dy \gg 1$, arising from real and virtual 
soft gluon emission compensate the strong coupling in the perturbation
series. These logarithms can be resummed using the techniques
of Balitsky, Fadin, Kuraev and Lipatov (BFKL) \cite{bfkl}.
Indeed it was first pointed out by Mueller and Navelet \cite{muenav}
that dijet production in high-energy hadron-hadron collisions would
be a particularly clean environment in which to look for evidence
of such resummation. They showed that the subprocess cross section
was expected to increase at asymptotic separations according
to $\hat{\sigma} \sim \exp{(\lambda \Delta y)}$ with  $\lambda = 
\alpha_s 12 \ln 2/\pi$. These ideas were taken further in 
Refs.~\cite{ds1,wjs,ds2,ds3}. In particular, it was shown that
the {\em azimuthal decorrelation} of the jet pair, resulting
from multiple soft gluon emission in the rapidity interval between
them, provides a particularly distinctive signature of BFKL
dynamics, free of potentially confounding effects from dependence
on parton distributions.
The leading BFKL resummation provides a quantitative
prediction for the rate of decorrelation with increasing separation
$\dy$.  

The azimuthal decorrelation has recently been measured by the 
D0 collaboration \cite{dzeropl} at the Tevatron $\pp$ collider.
The results are intriguing. The observed rate of decorrelation is
larger than that predicted by an `exact' next-to-leading order
calculation based on the $2\to 2$ and $2\to 3$ matrix elements (as
implemented in the JETRAD program \cite{JETRAD}), but smaller than
that predicted by both the leading BFKL resummation \cite{ds1,wjs}
and an `improved' BFKL calculation incorporating certain
subleading kinematical effects \cite{ds2}. In fact the data agree
best with the predictions of the HERWIG parton-shower Monte Carlo
\cite{HERWIG}, based on DGLAP multigluon emission with angular ordering.

The bulk of the theoretical attention concerning BFKL dynamics 
has focused on its application to HERA ep collider physics -- originally,
to the behavior of the 
structure function $F_2$ at small $x$, and latterly to more
exclusive quantities (for a review see Refs.~\cite{jkreview,vddreview} and
references therein). In particular, `forward jet' production has been
studied in Refs.~\cite{kms94,agkm,bddg,mz,bdw},
and forward single particle and single
photon production in Ref.~\cite{klangm}.
The possibility of observing
`BFKL gluons' in small-$x$ deep inelastic scattering has been discussed
in Ref.~\cite{klm}.
One result of this activity is the realization
that subleading corrections to the leading BFKL resummation,
from effects such as the running coupling and phase space, are likely
to be numerically important in practice; see for example Ref.~\cite{kms}.
Unfortunately the complete perturbative next-to-leading logarithmic corrections
are not yet available in a form which allows them to be incorporated
in a phenomenological analysis, although there has been considerable
theoretical progress towards this goal \cite{NLLfl,NLLvdd,NLLcc}.

The aim of the present work is to study such subleading effects and 
to  see whether they
could account for the discrepancy between the BFKL predictions and the
Tevatron data. We do this by recasting the original analytic resummed
expression for the parton cross section \cite{muenav} in the form
of an event generator. 
This not only allows kinematic subasymptotic constraints
(energy conservation, for example) 
and other corrections to be readily implemented, but also gives a better
overall picture of the events with large dijet rapidity separation
(for example, the distribution  of the accompanying BFKL
`minijets'). A further advantage is that experimental acceptance cuts
on the jets are straightforward to impose.  
A similar Monte Carlo approach was used recently in the fixed coupling limit 
\cite{schmidt} to look at transverse
energy flow in dijet production.

In this paper we concentrate on several issues. First, we set
up the calculational framework and demonstrate  that we are able  to reproduce
the results based on the analytic resummed expressions of 
Refs.~\cite{muenav,ds1,wjs}. Second, we examine the effect on the 
predicted azimuthal decorrelation  of including various subasymptotic
effects, including kinematic constraints and the running coupling constant.
We will show that these effects are indeed important in the kinematic
domain accessible to the Tevatron experiments.  More general applications
of our formalism, 
together with a more detailed comparison with data, will be presented
elsewhere.

\section{BFKL formalism for dijet production}

\subsection{Summary of analytic results}

We wish to describe events in hadron collisions 
containing two jets with relatively small
transverse momenta $\pti, \ptii$  and large rapidity separation
$\dy\equiv y_1 - y_2$.  In the limit where the minimum jet transverse
momentum $P_T$ is small compared to the parton-parton center-of-mass energy
and $\dy$ is large,
the inclusive dijet cross section can be written analytically
to leading logarithmic order in the BFKL 
approach \cite{muenav}.   

For jets produced in gluon-gluon collisions
(the $q\bar q$ initial state 
can be included via the effective subprocess approximation;
see below) the differential cross section is given by 
\beq
{d \hat{\sigma}_{gg} \over d \ptisq d \ptiisq d\df}
 = {\alpha_s^2 C_A^2 \pi\over 2 p_{T1}^3 p_{T2}^3 } \;
\frac{1}{2\pi} \sum_{n=-\infty}^{+\infty} {\rm e}^{i n \df}\;
\frac{1}{2\pi} \Re \int_{-\infty}^{+\infty} dz
\; \exp{\left(2t \chi_n(z) + iz\ln(\ptisq/ \ptiisq)\right)}
\label{eq:a1}
\eeq
with $t = \alpha_s C_A \dy /\pi$  and
\beq
\chi_n(z) = \Re\left[ \psi(1) - \psi\left(\half(1+\vert n\vert)
+iz\right) \right] \; .
\label{eq:a2}
\eeq
Here $\psi$ is the logarithmic derivative of the gamma function.
We have defined
\beq
\df\equiv \vert \phi_1-\phi_2 \vert -\pi
\eeq
so that $\df=0$ when the two jets are back-to-back in the transverse plane.

Integrating over the dijet transverse momenta above a fixed
threshold $P_T$ then gives for the azimuthal distribution
\beq
{d \hat{\sigma}_{gg} \over  d\df}\Bigg\vert_{\ptisq,\ptiisq > P_T^2}
 = {\alpha_s^2 C_A^2 \pi\over  2 P_T^2} \;
\frac{1}{2\pi} \sum_{n=-\infty}^{+\infty} {\rm e}^{in\df} C_n(t)\; , 
\label{eq:a3}
\eeq
with
\beq
C_n(t)  = \frac{1}{2\pi}  \int_{-\infty}^{+\infty}
{ dz \over z^2 + \quarter}
\;  \exp{\left(2t \chi_n(z)\right)}  \; .
\label{eq:a4}
\eeq
The total subprocess cross section is simply proportional to
$C_0(t)$:
\beq
 \hat{\sigma}_{gg}  =  {\alpha_s^2 C_A^2 \pi\over  2 P_T^2}\;
 C_0(t).
\label{eq:a5}
\eeq
Its asymptotic behavior is determined by
\beq
% \hat{\sigma}_{gg}  =  {\alpha_s^2 C_A^2 \pi\over  2 P_T^2}\;
 C_0(t)\ \left\{
\begin{array}{ll}
= 1 & \mbox{for} \ \   t = 0 \\
  \sim  
\left[ \half \pi 7 \zeta(3) t \right]^{-1/2} \; {\rm e}^{4 \ln 2\; t}\;
 & \mbox{for} \ \   t \to \infty
\end{array} \right.
\label{eq:a6}
\eeq
from which we see the characteristic BFKL prediction of an exponential
increase in the cross section with large $\dy$ (equivalently, large $t$).
It can also be seen from (\ref{eq:a3}) that the average  
cosine of  the  azimuthal angle 
difference $\df$ defined above is proportional to $C_1(t)$.  In fact we have
\beq
\langle \cos\df \rangle= {{C_1(t)}\over{C_0(t)}}
\label{eq:a7}
\eeq
and as we shall see below, this falls off with increasing $t$,
exhibiting the decorrelation expected with the emission
of gluons in the rapidity interval between the jets.

In what follows we will compare the analytic results just described with 
results from our formalism described in the next subsection.  
We will pay particular
attention to the total subprocess cross section (\ref{eq:a5})
and the azimuthal angle correlation (\ref{eq:a7}).

\subsection{Iterated solution for use in event generators}

The assumptions built into the leading order BFKL formalism that
allow us to obtain an analytic solution for the cross section 
are not all easily satisfied in an experimental situation.
In particular, implicit in the above solution are integrations
over the transverse momenta of intermediate radiated gluons that 
extend to infinity.  Furthermore nonleading effects (which 
lead for example to the running of $\alpha_s$) are also neglected.
And finally, the analytic solution (\ref{eq:a3},\ref{eq:a4})
is symmetric with respect to the two
observed jets, while  experimental cuts may not be. 

In this section we solve the BFKL equation by iteration, which 
allows us to avoid these assumptions and obtain a solution that is 
more directly amenable to comparison with experiment.  This 
solution amounts to `unfolding' the summation over the intermediate radiated
gluons and making their contributions explicit.  It is then straightforward
to implement this iterated solution in an event generator, as
we describe below.

To obtain the iterated solution we begin with the differential cross section
\beq
{d \hat{\sigma}_{gg} \over d^2 p_{T1}d^2 p_{T2} d \dy}
 = {\alpha_s^2  C_A^2 \over \ptisq \ptiisq }\;
   f(\vpti, \vptii, \dy).
\label{eq:sighat}
\eeq
The Laplace transform $\tf$ of the function $f$ with respect to $\dy$
satisfies the BFKL equation.  Defining 
\beq
\tf(\vpti,\vptii, \omega) = \int_0^\infty \d\Delta 
\; e^{-\omega\Delta}\; f(\vpti,\vptii, \Delta) \; ,
\label{eq:b2}
\eeq
where for convenience we use $\Delta\equiv\Delta y$ ,
we have
\bea
\omega \tf(\vpti,\vptii, \omega) &=& 
\delta(p_{T1}^2 -p_{T2}^2)\delta(\phi_1-\phi_2) 
+\left({\as C_A \over \pi^2} \right)   \nonumber \\
&&\times \int{ \d^2q_T\over q_T^2} 
\left[ \tf(\vpti+\vec{q_{T}},\vptii, \omega)
 - {p_{T1}^2\tf(\vpti,\vptii, \omega)
\over q_T^2 + (\vpti+\vec{q_T})^2 }\right] .
\label{eq:bfkl}
\eea
which is the BFKL equation for dijet production in
hadron collisions.  Infrared divergences from real gluon emission -- 
the first term in the integral on the right-hand side -- are cancelled
by the virtual gluon contribution in the second term.  Note that 
in writing Eq.~(\ref{eq:bfkl}) we have assumed that $\as$ is fixed.

Instead of solving (\ref{eq:bfkl}) analytically, which would lead to the results 
shown above, we solve iteratively, using a slightly modified form of
the equation.   
Following Ref.~\cite{klm}, we note that very low 
energy gluons are not resolvable and we therefore separate the real gluon
integral into `resolved' and `unresolved' contributions,
according to whether they lie above or below a small transverse energy
scale $\mu$.  The scale $\mu$ is assumed to be
small compared to the other relevant scales in the problem (the minimum
transverse momentum $P_T$ of the `external' jets, for example).  
We then combine the virtual and unresolved 
contributions into a single, finite integral.  The BFKL
equation then becomes 
\bea
\omega \tf(\vpti,\vptii, \omega) &=& 
\delta(p_{T1}^2 -p_{T2}^2)\delta(\phi_1-\phi_2) 
+\left({\as C_A \over \pi^2} \right)  
\int_{q_T^2>\mu^2}{ \d^2q_T\over q_T^2} 
\tf(\vpti+\vec{q_{T}},\vptii, \omega) \nonumber \\
&&- \left({\as C_A \over \pi^2} \right)
\int{ \d^2q_T\over q_T^2} 
\Biggl[ \tf(\vpti+\vec{q_{T}},\vptii, \omega)\theta(\mu^2-q_T^2)
 \nonumber \\
&& 
- {p_{T1}^2\tf(\vpti,\vptii, \omega)
\over q_T^2 + (\vpti+\vec{q_T})^2 }\Biggr]  .
\label{eq:b4}
\eea
The combined unresolved/virtual integral can be simplified by noting 
that $q_T \ll \pti$ by construction  because $\mu^2 \ll P_T^2$ and $\pti>P_T$.
Therefore 
\beq
\tf(\vpti+\vec{q_T},\vptii,\omega)\approx\tf(\vpti,\vptii,\omega),
\label{eq:approx}
\eeq
which allows us to write 
\bea
(\omega-\omega_0) \tf(\vpti,\vptii, \omega) &=&
\delta(p_{T1}^2 -p_{T2}^2)\delta(\phi_1-\phi_2) \nonumber \\
&&+\left({\as C_A \over \pi^2} \right)
\int_{q_T^2>\mu^2}{ \d^2q_T\over q_T^2}
\tf(\vpti+\vec{q_{T}},\vptii, \omega),
\label{eq:bfklfinal}
\eea
where we have defined
\beq
\omega_0 \equiv\left({\as C_A \over \pi^2} \right) 
\int{ \d^2q_T\over q_T^2} 
\left[ \theta(\mu^2-q_T^2)
 - {p_{T1}^2
\over q_T^2 + (\vpti+\vec{q_T})^2 }\right] .
\label{eq:omega0}
\eeq
The virtual and unresolved contributions are now contained
in $\omega_0$ and we are left with an integral over resolved real
gluons.

We now solve Eq.~(\ref{eq:b6}) iteratively, and performing the inverse transform
we have 
\beq
 f(\vpti, \vptii, \Delta ) = 
  \sum_{n=0}^{\infty} f^{(n)}(\vpti, \vptii, \Delta )
\label{eq:b7}
\eeq
where the exact form of $f^{(n)}$ depends on whether the coupling $\alpha_s$
is taken to be fixed or running.  Strictly speaking, the derivation above
only applies for fixed coupling because we have left $\alpha_s$
outside the integrals.  The modifications necessary to account for 
running coupling are discussed below.

\subsubsection{Solution for fixed $\as$}

For the case where $\alpha_s$ is fixed 
(evaluated at the scale $P_T^2$, for example),
the integral in (\ref{eq:b6}) is straightforward to evaluate and 
we obtain 
\beq
\omega_0 = \left({\as C_A \over \pi^2} \right) 
    \ln(\mu^2/p_{T1}^2).
\eeq
Substituting into (\ref{eq:bfklfinal}), solving by iteration and then 
performing the inverse Laplace transform, we find
\beq
f^{(0)}(\vpti,\vptii,\Delta)  = 
\left[ {\mu^2\over \ptisq}\right]^{\as C_A\Delta/\pi}\
\delta^{(2)}(\vpti+\vptii )
\label{eq:fzero}
\eeq
and for $n\neq 0$
\bea
f^{(n)}(\vpti,\vptii,\Delta) & = &
\left( {\as C_A\over\pi^2} \right)^n
\left[ {\mu^2\over\ptisq  }\right]^{\as C_A\Delta/\pi }\
\prod_{i=1}^{n}  \int {d^2 q_{Ti}\over{q_{Ti}^2 }}\;  
\theta(q_{Ti}^2 - \mu^2)\; \nonumber \\
& & \times \int_0^\Delta d Y_1 \int_0^{Y_1}d Y_2 \ldots \int_0^{Y_{n-1}}
d Y_n \;
\delta^{(2)}(\vpti+\vptii + \sum_{i=1}^{n}\vec{q}_{Ti} ) \nonumber \\
&&\times \left[ {\ptisq
       \over (\vpti+\vec{q}_{T1})^2 }\right]^{\as C_A Y_1/\pi }
\left[ {(\vpti+\vec{q}_{T1})^2 ) 
   \over (\vpti+\vec{q}_{T1}+\vec{q}_{T2})^2 }\right]^{\as C_A Y_2/\pi }
\nonumber \\
& & \qquad \ldots
\left[ {(\vpti+\sum_{i=1}^{n-1}\vec{q}_{Ti})^2
     \over (\vpti+\sum_{i=1}^{n}\vec{q}_{Ti})^2 }\right]^{\as C_AY_n/\pi }.
\label{eq:fn}
\eea
The inverse transform has given rise to a set of nested integrals
over the variables $Y_i$, which can be interpreted as the rapidities of the
emitted gluons.
The differential subprocess cross section is then given by
\beq
{d \hat{\sigma}_{gg} \over d^2 p_{T1}d^2 p_{T2} d \dy}
 = {\alpha_s^2  C_A^2 \over \ptisq \ptiisq }\;
  \sum_{n=0}^{\infty} f^{(n)}(\vpti, \vptii, \Delta ).
\label{eq:b11}
\eeq
The subprocess cross section is now expressed as an explicit sum
over radiated gluons, with corresponding $\vec{q_{Ti}}$ and $Y_i$
integrals over their phase space.  It is straightforward to implement 
this in a Monte Carlo event generator,\footnote{See also Ref.~\cite{schmidt}
for a similar approach.}
 and to impose energy conservation
and experimental cuts via limits on the integration; see below
for numerical results.

An interesting feature of the results (\ref{eq:fzero}) and 
(\ref{eq:fn}) is the presence of the form factors 
$$\left[ {\mu^2\over\ptisq  }\right]^{\as C_A\Delta/\pi }
\quad \mbox{and} \quad 
\left[ {(\vpti+\sum_{i=1}^{n-1}\vec{q}_{Ti})^2
     \over (\vpti+\sum_{i=1}^{n}\vec{q}_{Ti})^2 }\right]^{\as C_AY_n/\pi } ,$$
respectively.  These form factors arise  
from the resummation of the unresolved
$q_{Ti}^2 < \mu^2$ soft gluon emission in the rapidity interval $\Delta$.
In particular we see the modification of the 
naive zeroth-order perturbative result $\delta(\vpti + \vptii)$
by 
\beq
\left[ {\mu^2\over \ptisq }\right]^{\as C_A\Delta/\pi}
< 1 \qquad \mbox{for} \ \   \ptisq > \mu^2, \ \Delta > 0.
\label{eq:b6}
\eeq
This is a consequence of the fact that the emission of soft gluons 
reduces the probability of the dijets having equal transverse momenta and 
 being back-to-back in
azimuth. For $\Delta = 0$ all radiation is suppressed (in this
approximation) and the form factor is equal to unity.

We can make contact with the analytic results of the previous section 
by noting that the only additional approximation we have made is 
(\ref{eq:approx}), which we used in the computation of $\omega_0$, 
thereby neglecting momenta smaller than $\mu$ compared
to $p_{T1}$.  We therefore expect our result to agree with the 
analytic one up to corrections of $\cO(\mu^2/p_{Ti}^2)$.
In fact one can show that the 
dijet cross section (\ref{eq:b11}) integrated over the external transverse
momenta
\beq
{d \hat{\sigma}_{gg} \over  d\df} =
\int d^2 p_{T1}d^2 p_{T2}\; \delta(\vert \phi_1-\phi_2 \vert -\pi -\df)\;
\theta(\ptisq-P_T^2)\theta(\ptiisq-P_T^2)\;
 {d \hat{\sigma}_{gg} \over d^2 p_{T1}d^2 p_{T2} d \dy}
\label{eq:b5}
\eeq
corresponds {\it exactly} to the analytic result (\ref{eq:a5}) in the
limit $\mu^2 \to 0$. This will be illustrated numerically below.
In the simulations based on Tevatron kinematics to be described in the
following section, where $p_{Ti} > \cO(10\; \GeV)$, we will use
values $\mu = \cO(1\; \GeV)$ such that the finite $\cO(\mu^2/p_{Ti}^2)$
corrections are negligible.

\subsubsection{Solution for running $\as$}

Higher order corrections are known to lead to the running of 
the coupling constant $\as$ \cite{NLLcc}.  
They can therefore be taken into
account by including momentum dependence in the coupling
associated with the emission of each gluon.
We shall see that this leads to two types of modification:
(i) the factors of $\as$ associated with resolved real gluon emission are
simply evaluated at the scale of the emitted gluon; and
(ii) the form factors associated with the unresolved real and 
virtual gluon emission get modified slightly.

We will include the running to
lowest order, taking 
\beq
\as(q^2) = {1\over b\ln(q^2/\Lambda^2)}
\label{eq:as}
\eeq
where
\beq
      b = { 33-2N_f \over  12 \pi}
\label{eq:b}
\eeq
with $N_f=4$.
We then pull the 
factors of $\alpha_s$ in (\ref{eq:bfklfinal}) and 
(\ref{eq:omega0}) inside the integrals and make the substitution 
$\as \rightarrow \as(q_T^2)$.
In addition, we must regulate the behavior of $\as(q^2)$
to prevent its becoming unphysically large   
as $q^2$ becomes small.  This can happen for 
example in the momentum integration in the expression for  $\omega_0$.  
Here we simply assume that the value of $\alpha_s$ freezes out 
below some scale $Q_0 > \Lambda$, i.e., we take
\beq
\as(q^2)\ \left\{
\begin{array}{ll}
= \asnot\equiv\as(Q_0^2) & \mbox{for} \ \   q^2 \leq Q_0^2\\
  = \as(q^2)  
 & \mbox{for} \ \   q^2 \ge Q_0^2
\end{array} \right.
\label{eq:asfreeze}
\eeq
In practice choosing values  $\Lambda < Q_0  < \mu$ 
so that $\asnot$ is of ${\cal O}(1)$, as we do below, 
gives results that are insensitive to the exact choice.

The iterated solution to the BFKL equation in the running coupling case
is then given by 
\bea
f^{(n)}(\vpti,\vptii,\Delta) & = &
\left[ {\as(\ptisq)\over \as(\mu^2) }\right]^{C_A\Delta/\pi b}\
\prod_{i=1}^{n}  \int d^2 q_{Ti}\;  \theta(q_{Ti}^2 - \mu^2)\;
{\as(q_{Ti}^2) C_A \over \pi^2 q_{Ti}^2 }\; \nonumber \\
& & \times \int_0^\Delta d Y_1 \int_0^{Y_1}d Y_2 \ldots \int_0^{Y_{n-1}}
d Y_n \;
\delta^{(2)}(\vpti+\vptii + \sum_{i=1}^{n}\vec{q}_{Ti} ) \nonumber \\
&&\times \left[ {\asb((\vpti+\vec{q}_{T1})^2)
       \over \asb(\ptisq ) }\right]^{C_A Y_1/\pi b}
\left[ {\asb((\vpti+\vec{q}_{T1}+\vec{q}_{T2})^2)
       \over \asb((\vpti+\vec{q}_{T1})^2 ) }\right]^{C_A Y_2/\pi b}
\nonumber \\
& & \qquad \ldots
\left[ {\asb((\vpti+\sum_{i=1}^{n}\vec{q}_{Ti})^2)
     \over \asb((\vpti+\sum_{i=1}^{n-1}\vec{q}_{Ti})^2 ) }\right]^{C_AY_n/\pi b}
\label{eq:fnrunning}
\\
f^{(0)}(\vpti,\vptii,\Delta) & = &
\left[ {\as(\ptisq)\over \as(\mu^2) }\right]^{C_A\Delta/\pi b}\
\delta^{(2)}(\vpti+\vptii )
\label{eq:fzerorun}
\eea
where
\beq
\asb(q^2)\ \left\{
\begin{array}{ll}
=  (Q_0^2/q^2)^{\asnot b} \asnot & \mbox{for} \ \   q^2 \leq Q_0^2\\
  = \as(q^2)  
 & \mbox{for} \ \   q^2 \ge Q_0^2
\end{array} \right.
\label{eq:asbar}
\eeq
Thus the result for running $\as$ is obtained by the replacements 
\bea
\alpha_s^n & \to & \prod_{i=1}^{n}\overline{\alpha}_s(q_{Ti}^2) \nonumber \\
\left[ {\mu^2 \over q^2 }\right]^{\alpha_s C_A Y /\pi}
&  \to &
\left[ {\asb(q^2)\over \asb(\mu^2) }\right]^{C_A Y /\pi b}
\label{eq:replace}
\eea
in the fixed coupling results (\ref{eq:fzero},\ref{eq:fn}) for $f^{(0)}$ and 
$f^{(n)}$ and
\beq
\as^2\to\as(p_{T1}^2) \as(p_{T2}^2)
\eeq
in the differential cross section (\ref{eq:b11}).
Note that $\asb$ is equal to $\as$ unless it is evaluated at a momentum
smaller than the `freeze-out' scale $Q_0$.  We maintain the hierarchy of
scales $\Lambda<Q_0<\mu\ll P_T$.

\subsection{Cross section}

The calculation is completed by weighting the integrand in the definition
of $f^{(n)}$ (Eqs.~(\ref{eq:fn},\ref{eq:fnrunning})) 
in the subprocess cross section with parton distributions
$G(x_1, Q^2)  G(x_2, Q^2)$ where, 
 using
the `effective subprocess approximation',
\beq
G(x, Q^2)  =   g(x,Q^2)\;  +\;
 {4 \over 9}\; \sum_{q=u,d,s,c} (q(x,Q^2) + \bar{q}(x, Q^2) ) \; .  
 \label{eq:partons}
\eeq
The parton momentum fractions $x_1$ and $x_2$ are determined
by the  invariant mass $\sqrt{\hat{s}}$ and rapidity $Y$ 
of the multijet final state:
\bea
x_1 & = & \sqrt{{\hat{s}\over s}}\;  {\rm e}^Y 
\; = \; 
{ 1 \over \sqrt{s} }\; {\rm e}^{-\Delta/2 + Y} \; \left(  
p_{T1}\; {\rm e}^\Delta 
\;+\; p_{T2}
\; + \; \sum_i q_{Ti} \; {\rm e}^{Y_i}
\right) 
\nonumber \\
x_2 & = &\sqrt{{\hat{s}\over s}} \; {\rm e}^{-Y}
\; = \; 
{ 1 \over \sqrt{s} } \; {\rm e}^{\Delta/2 - Y}  \;  \left(  
p_{T1}\; {\rm e}^{-\Delta} 
\;+\; p_{T2}
\; + \; \sum_i q_{Ti} \; {\rm e}^{-Y_i}
\right) 
\label{eq:x1x2}
\eea
where $\sqrt{s}$ is the total collision energy. In the numerical studies
to be described below we choose the factorization scale $Q^2 = P_T^2$.

Note that the requirement $x_1,x_2 \leq 1$ effectively imposes
an upper limit on the transverse momentum ($q_{Ti}$) integrals.
This in turn means that the analytic results (\ref{eq:a2},\ref{eq:a3})
are {\it not} reproduced in the presence of such a constraint,
since they require the internal transverse momenta integrals
to extend to infinity. In the original work of Ref.~\cite{muenav}
(see also \cite{wjs}) the strong ordering of the rapidities was used
to approximate the right-hand side of (\ref{eq:x1x2}) by 
\bea
x_1 & = &
 { p_{T1}\over \sqrt{s}} \;  {\rm e}^{\Delta/2 + Y}
\nonumber \\
x_2 & = &
 { p_{T2}\over \sqrt{s}} \;  {\rm e}^{\Delta/2 - Y}
\label{eq:x1x2approx}
\eea
 so that there was no longer any constraint
on the $q_{Ti}$. An improved approximation where account was taken
of the additional energy required for the multigluon emission was
studied in Ref.~\cite{ds2}.  In our approach all kinematic constraints
are applied directly to the multijet final state.

\section{Numerical Results}

In this section we present numerical results from our event generator for 
dijet production in the BFKL approach.  We will take a brief
look at the properties of the subprocess cross section $\hat\sigma$
and then present predictions for the Tevatron $\pp$ collider with 
center-of-mass energy 1.8 TeV.

In our numerical computations we make the following choices for the
relevant parameters.  The minimum transverse momentum for each jet
of the pair is $P_T=20\ \GeV$, and for simplicity the two
jets are assumed to have equal and opposite rapidities:
$y_1=-y_2=\Delta/2$.  Unless otherwise noted 
the scale $\mu$ that defines the 
boundary between resolved and unresolved gluons we take to be $\mu=1\ \GeV$.
In the fixed coupling case we evaluate $\alpha_s$ at the scale $P_T^2$;
for $\Lambda=200\ \MeV$ (as dictated by our choice of parton 
distributions \cite{grv}; see below) and our choice of $P_T$, this 
gives $\as=0.164$.
In the running coupling case we again have $\Lambda=200\ \MeV$ and we choose
$Q_0=0.425\ \GeV$ so that $\asnot=1$.
Finally, in our Tevatron calculations we use the leading-order parton
distribution functions of Ref.~\cite{grv}.

We begin with the subprocess cross section, shown in Figure 1(a) as a 
function of the dijet rapidity difference $\Delta$, normalized to 
its value at $\Delta=0$.  The results for fixed $\as$ are shown as
open circles (the error bars shown are from Monte Carlo statistics), 
and we see the characteristic BFKL exponential rise 
with increasing $\Delta$.  We also see that the analytic result, 
shown as a solid curve, is well-reproduced by the iterated solution.
The cross section for running $\as$ is shown as the points marked
`x', and we see that the running of the coupling has the effect 
of slightly suppressing the increase in $\hat\sigma$ with $\Delta$ compared
to the fixed case.  
In Figure 1(b) we show $\langle \cos\df \rangle$ computed from the
subprocess cross section $\hat\sigma$ for the same cases as 
in Fig.~1(a).  The fixed-$\as$ results (circles) agree very well
with the analytic prediction (solid curve), while the running $\as$
results show a slightly slower decorrelation.  We will see a further
slowing of the decorrelation when conservation of energy and 
parton distributions are included below.

In Figure 2 we investigate the sensitivity of our results to the choice of 
gluon resolution cutoff $\mu$.  As noted above, we expect corrections
to be of order $\mu^2/P_T^2$ so that the $\mu$ dependence becomes
stronger as $\mu$ approaches $P_T$, but for $\mu$ small enough there should
be relatively little sensitivity.  This is illustrated in Fig.~2 where
we show the cross section $\hat\sigma$ as a function of $\mu$ for 
two values of the dijet rapidity difference $\Delta$.
In this and the following figures we show results for running 
$\as$.  We see
the expected behavior and also that larger values of $\Delta$
exhibit more sensitivity to $\mu$.  This is related to the fact, as we will
see explicitly below, that larger $\Delta$ means more emitted gluons.
This in turn leads to more sensitivity to the resolution parameter.
We also note that the presence of other scales in 
the problem besides $P_T$ 
($\Lambda, Q_0$) mean that arguments about the size of corrections
are not rigorous.
In any case, it is clear from the figure that our choice of 
$\mu=1\ \GeV$ is sufficient to guarantee that we are not senstive
to its exact value.

We examine in Figure 3 how the cross section is distributed amongst
the contributions from different numbers of emitted gluons for
the same values of $\Delta$ as in the previous figure.  As
expected, larger rapidity differences between the jets allow for larger
numbers of emitted gluons, so that the cross section peaks
at higher $n_g$ with increasing $\Delta$.  Note that the increasing 
area of the histograms simply reflects the rise of the cross section
with increasing $\Delta$.  We should point out that there is {\it not} a 
direct correspondence between  emitted
gluons and physical jets, so that these results for $n_g$ should
not be interpreted directly as a prediction for numbers of jets.
By the same token, we note that the distribution in $n_g$ depends 
to some extent on the resolution cutoff parameter $\mu$
  --- larger values of $\mu$ lead to fewer resolvable gluons.

We now turn from the subprocess cross section to the total cross section
integrated over parton distributions.  Figure 4 shows the cross section
as a function of dijet rapidity difference $\Delta$ for fixed and running
$\as$.  In both cases the falling parton densities more than compensate for the 
rise in $\hat\sigma$ with the  net result that the cross section falls off
with increasing $\Delta$.  
%Comparing the BFKL results with lowest-order
%QCD, shown as a solid curve in the figure, shows that the BFKL cross sections
%fall off more slowly than lowest-order QCD.  The latter curve simply 
%reflects the reduction in available phase space as $\Delta$ increases.  
The slower rise in $\hat\sigma$
for the running case that we saw in Fig.~1 translates here into a 
faster fall-off than one expects in the fixed coupling case.  
The lowest-order QCD result is shown as a solid curve.  Its slower
fall-off than the two BFKL curves reflects the fact that it includes 
lowest-order kinematics only, whereas the BFKL results include kinematic
suppression due to emission of extra gluons.

Because of the subtleties involved in untangling the BFKL prediction for 
$\hat\sigma$ from the effects of the falling parton densities in the 
measured dijet cross section, it was proposed in Refs.~\cite{ds1,wjs}
to measure the azimuthal angle decorrelation between the two jets.
This quantity is relatively insensitive to the details of the parton
densities and provides a clear distinction between the predictions of
next-to-leading-order QCD and BFKL.  As noted above, previous comparisons 
were to BFKL predictions which did not account for subleading effects such as 
the running of $\as$ and truncation of transverse momentum integrals
for the emitted gluons.  In Figure 5 we present our prediction for
$\langle \cos\Delta\phi\rangle$ with these effects included.
For comparison, the analytic BFKL prediction is shown as  a solid curve.
Clearly, subleading effects are quite substantial.  In particular we see that 
the azimuthal decorrelation occurs more slowly with increasing 
dijet rapidity difference than predicted by  the analytic BFKL result.
This can be understood partly in terms of phase space availability ---
for a given  transverse momentum threshold, dijets produced at
larger rapidities require more energy, leaving less  phase
space available for emission of gluons.  This effect -- not present in the 
analytic solution -- partly mitigates the increasing probability for 
emitting more gluons in the center-of-mass system.  Because the additional
emitted gluons are responsible for the decorrelation, the result of
including subleading effects is a reduction in azimuthal decorrelation
compared to the analytic BFKL solution, as seen in the figure.

Finally, we return to the question that originally motivated this work:  can
the inclusion of subleading effects improve the agreement between the
measured dijet decorrelation and that predicted by BFKL?  
%%%
We show for
reference in Figure 5 some recent preliminary measurements from the 
\Dzero\  collaboration \cite{dzeropl} with the same minimum
jet transverse momentum of 20 GeV.  The comparison should be taken 
as a rough guide only, because our predictions (and our kinematic cuts) 
are at the parton level, and we assume equal and opposite rapidities of the 
dijets.  The \Dzero\ measurements (and cuts) are  at the jet 
level, and  the net dijet rapidity is allowed to range between $\pm 0.5$.
The finite width of the jets will lead, for example, to 
$ \langle \cos\df \rangle \neq 1$ even at $\Delta=0$.
Furthermore, the results are not final, and the error bars in the figure
represent the  statistical and uncorrelated systematic
uncertainties only; an error band showing 
correlated jet energy scale  systematic uncertainties appear at the 
bottom of the figure. See Ref.~\cite{dzeropl} for a 
full explanation of the data. 
Having made those qualifications, we note a marked improvement in agreement
between data and the BFKL prediction when subleading effects
are included.  

\section{Conclusions}

We have presented the formalism and numerical results for dijet production
at hadron colliders in the BFKL approach, using an improved 
formalism incorporating an iterated solution (as described
for deep inelastic scattering in \cite{klm})
that unfolds the sum over emitted gluons that is implicit in the 
analytic solution to the BFKL equation.
We have cast the iterated solution in the form of an event generator. This 
allows us to incorporate subleading effects such as energy conservation and
other kinematic constraints as well as the running of the strong coupling 
constant, which are necessarily absent in the analytic approach.  
It also allows us to examine the properties of dijet events.   
We find that the subleading effects included in the improved formalism 
can be substantial, and in particular they lead to improved agreement
with measurements of the azimuthal decorrelation in dijet production at
the Fermilab Tevatron.  Further elaboration of our results for hadron 
colliders,  as well as applications of the formalism to forward jet production
in ep collisions, will appear in future work.

\vspace{0.5cm}
\noindent {\bf Acknowledgements} \\
Useful discussions with Vittorio Del Duca and Terry Heuring are acknowledged.
LHO is grateful to the UK PPARC for a Visiting Fellowship and to the Centre
for Particle Theory at the University of Durham for hospitality while
part of this work was being completed.
This work was supported in part by the U.S. Department of Energy,
under grant DE-FG02-91ER40685 and by the U.S. National Science Foundation,
under grant PHY-9600155.

%\newpage

\newpage
\vskip 1truecm
\section*{Figure Captions}
\begin{itemize}
\item [{[\figi]}]
(a) The subprocess cross section $\hat\sigma$ for dijet production
as a function of the dijet rapidity difference $\Delta$ for
$\as$ fixed (circles) and running (x's), for minimum 
jet transverse momentum $P_T=20\ \GeV$ and $\mu=1\ \GeV$.
  The analytic solution
is shown as a solid curve.  In each case the cross section is normalized
to its value at $\Delta=0$.  Errors are from Monte Carlo statistics.
(b) The mean of $\cos \df$ computed from the subprocess cross section,
for the same cases as in (a).

\item [{[\figii]}]
The  
subprocess cross section $\hat\sigma$ for dijet production
as a function of the gluon resolution cutoff $\mu$ for $\Delta=1$
(circles) and $\Delta=3$ (x's) 
and $P_T=20\ \GeV$, with running $\as$.

\item [{[\figiii]}]
The contributions to the subprocess cross section $\hat\sigma$
 for dijet production from different numbers
 of emitted (i.e. resolved) gluons $n_g$, for $\Delta=1,3,5$, running $\as$,
$P_T=20\ \GeV$ and $\mu=1\ \GeV$.

\item [{[\figiv]}]
The total cross section for dijet production at the Tevatron 
as a function of the dijet rapidity difference $\Delta$ for 
$\as$ fixed (circles) and running (x's), for minimum 
jet transverse momentum $P_T=20\ \GeV$ and $\mu=1\ \GeV$.
  The lowest order QCD result  is shown as a solid curve.

\item [{[\figv]}]
The azimuthal angle decorrelation in dijet production at the Tevatron 
as a function of dijet rapidity difference $\Delta$ for running $\as$
(x's), for minimum 
jet transverse momentum $P_T=20\ \GeV$ and $\mu=1\ \GeV$.  
The analytic BFKL solution is shown as a solid curve 
and a preliminary measurement from \Dzero\ \cite{dzeropl} is shown
as diamonds.  Error bars on \Dzero\ points represent statistical and 
uncorrelated systematic errors;  correlated jet energy scale systematic 
are shown as an error band.   See text for further
explanation.

\end{itemize}

\newpage

\begin{figure}
\vspace*{23cm}
\hspace*{-3.5cm}
\includegraphics{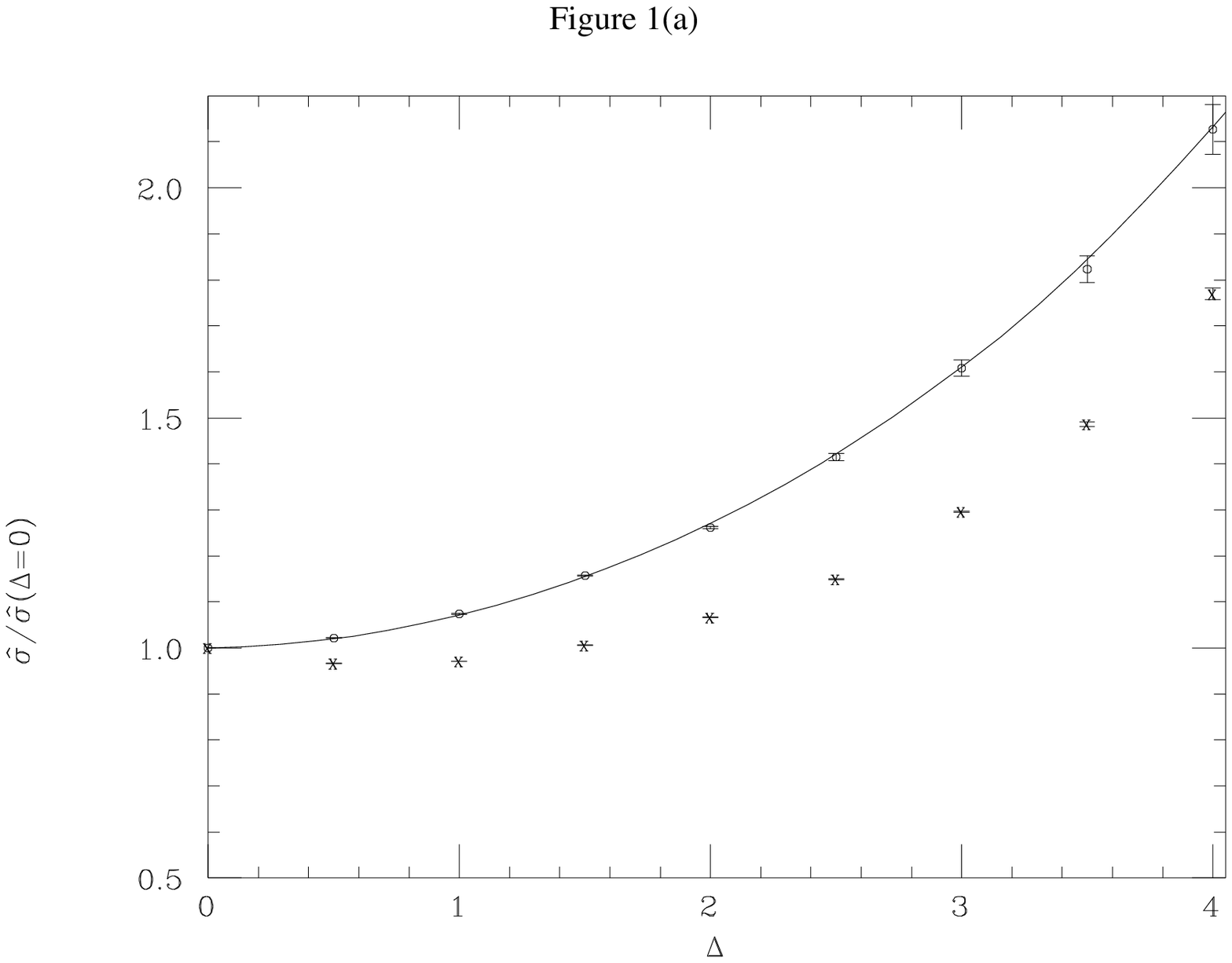}
\vspace{-7.5cm}
\end{figure}

\newpage

\begin{figure}
\vspace*{23cm}
\hspace*{-3.5cm}
\includegraphics{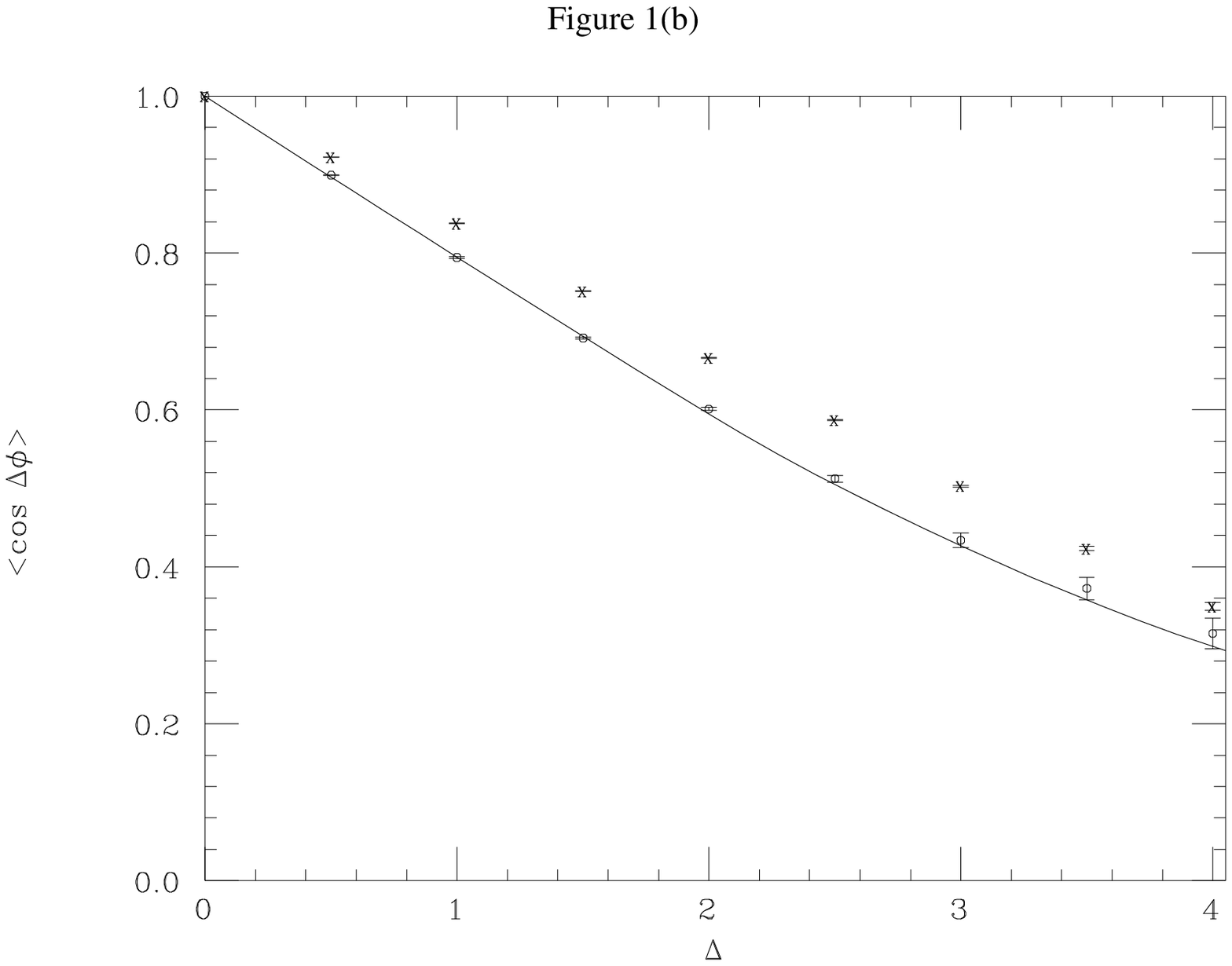}
\vspace{-7.5cm}
\end{figure}

\newpage

\begin{figure}
\vspace*{23cm}
\hspace*{-3.5cm}
\includegraphics{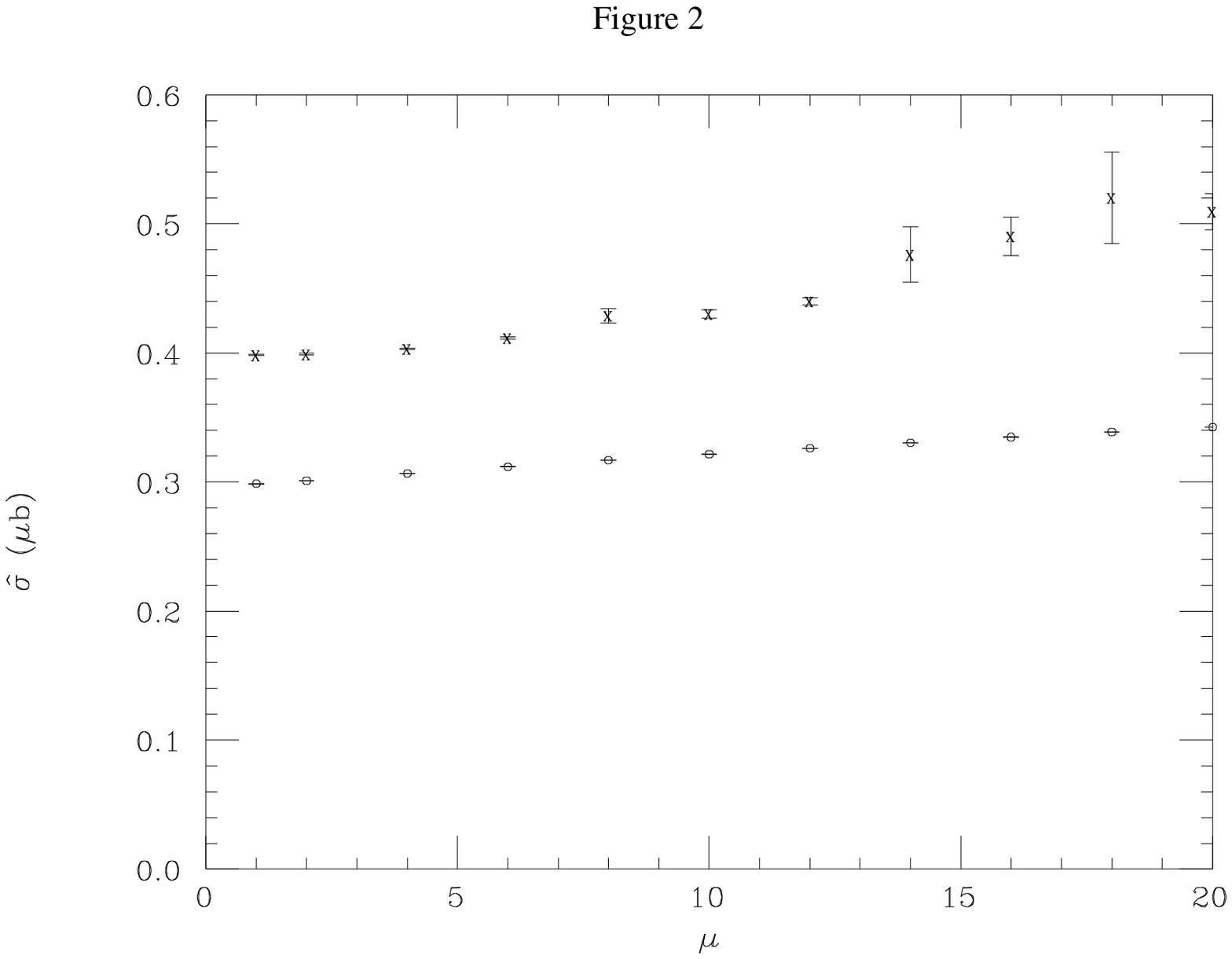}
\vspace{-7.5cm}
\end{figure}

\newpage

\begin{figure}
\vspace*{23cm}
\hspace*{-3.5cm}
\includegraphics{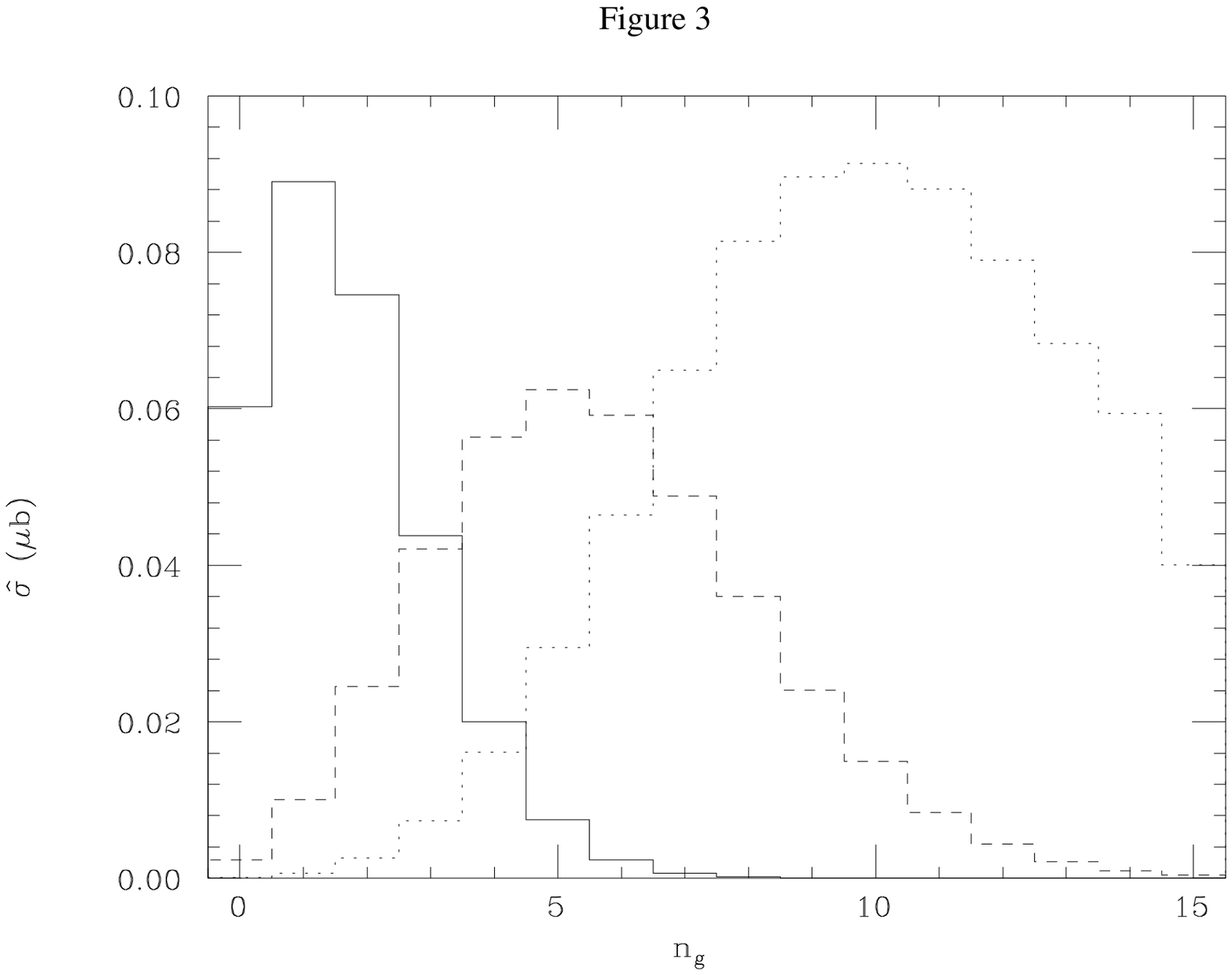}
\vspace{-7.5cm}
\end{figure}

\newpage

\begin{figure}
\vspace*{23cm}
\hspace*{-3.5cm}
\includegraphics{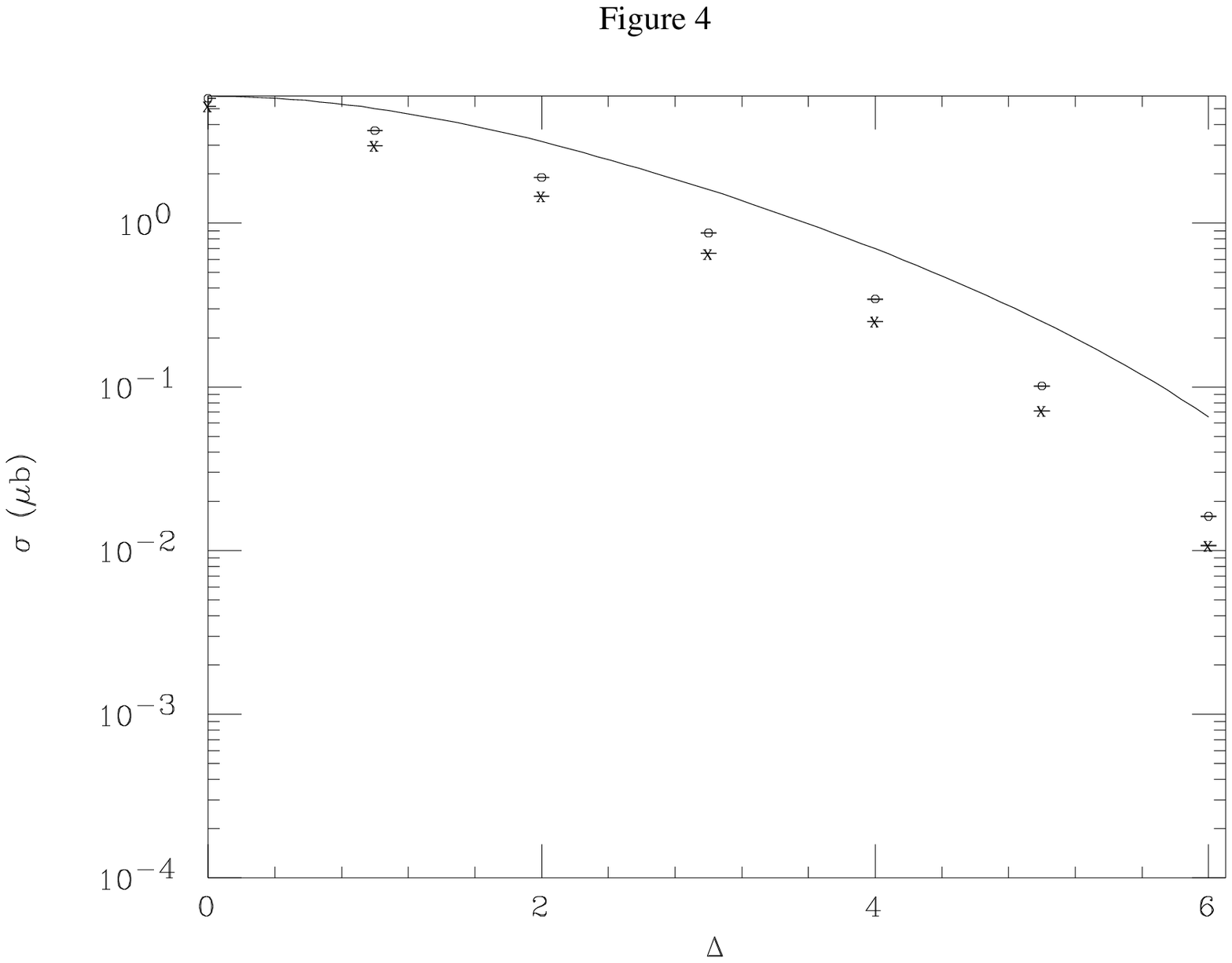}
\vspace{-7.5cm}
\end{figure}

\newpage

\begin{figure}
\vspace*{23cm}
\hspace*{-3.5cm}
\includegraphics{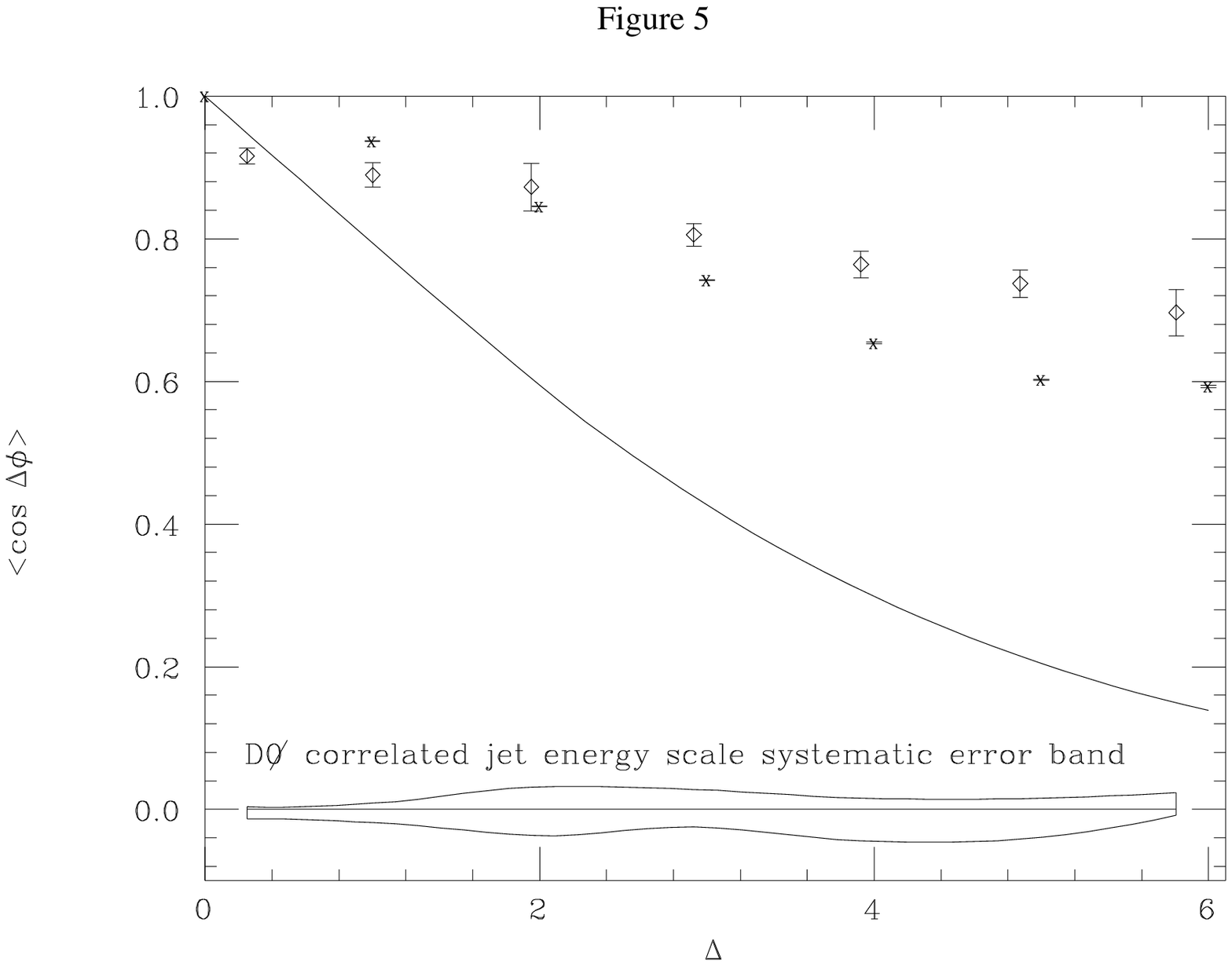}
\vspace{-7.5cm}
\end{figure}

\end{document}